\documentstyle[prb, preprint,dcolumn, epsfig, aps ]{revtex}
\usepackage{graphicx}

\begin{document}
\begin{frontmatter}
\title{Vortices and the mixed state of ultrathin Bi films}

\author{G. Sambandamurthy} 
\author{K. Das Gupta}
\author{Swati S. Soman}
\author{N. Chandrasekhar\thanksref{thank1}}
\address{ Department of Physics, Indian Institute of Science, 
Bangalore, India}

\thanks[thank1]{Corresponding author. 
 E-mail: chandra@physics.iisc.ernet.in}

\begin{abstract}
Current-voltage (I-V) characteristics of quench condensed, superconducting, ultrathin Bi films in a magnetic
field are reported. These I-V's show hysteresis for all films, grown both with and without thin Ge underlayers.
Films on Ge underlayers, close to superconductor-insulator transition, show a peak in the critical current,
indicating a structural transformation of the vortex solid. These underlayers, used to make the films more
homogeneous, are found to be more effective in pinning the vortices. The upper critical fields ($B_{\mathrm c2}$) of these
films are determined from the resistive transitions in perpendicular magnetic field. The temperature dependence
of the upper critical field is found to differ significantly from Ginzburg-Landau theory, after modifications for
disorder.
\end{abstract}

\end{frontmatter}

\section{Introduction}
Transport properties of disordered ultrathin films have
been studied extensively over the last decade especially in
the context of disorder driven and magnetic field driven
superconductor-insulator transition (SIT).~\cite{gra} This transition is
found to occur at a particular value of disorder corresponding
to a critical resistance Rc , which clusters close to $h/4e^2$
the quantum of resistance for Cooper pairs, for both field
driven and disorder driven transitions. Earlier studies on Bi,
quench condensed on underlayers of Ge, showed a ''homogeneous''
type of SIT which was characterized by a strong
suppression of the critical temperature Tc as disorder was
increased. The presence of a thin (10 ${\mathrm \AA}$) underlayer of Ge
is conventionally thought to improve the wetting properties
of the film, and thereby assist homogeneous growth.~\cite{str,nea} However,
several studies have indicated that the underlayer is not
inert, as conventionally assumed, and may actually play an
active role in determining the transport properties of the
films.

\section{Experimental Results}
Figures 1 and 2 show the I-V's on a semilog scale for Bi films
on bare quartz substrates, and on a Ge underlayer respectively,
until the normal state resistance is reached, where all the
curves at different magnetic fields meet. For the films on
bare substrates, only one critical current can be unambiguously
identified. For the films on Ge underlayers, two different
steplike increases in the voltage can be clearly identified
for three different field values. It is only after the second step
that the film resistance becomes the normal state resistance,
and therefore this should be defined as the critical current.
Hence the first step has to be associated with some sort of
structural transition in the vortex solid. Recall that the depinning
current is lower than this current at which the first step
in the voltage appears.

\begin{figure}[t]
\begin{center}\leavevmode
\includegraphics[width=0.8\linewidth]{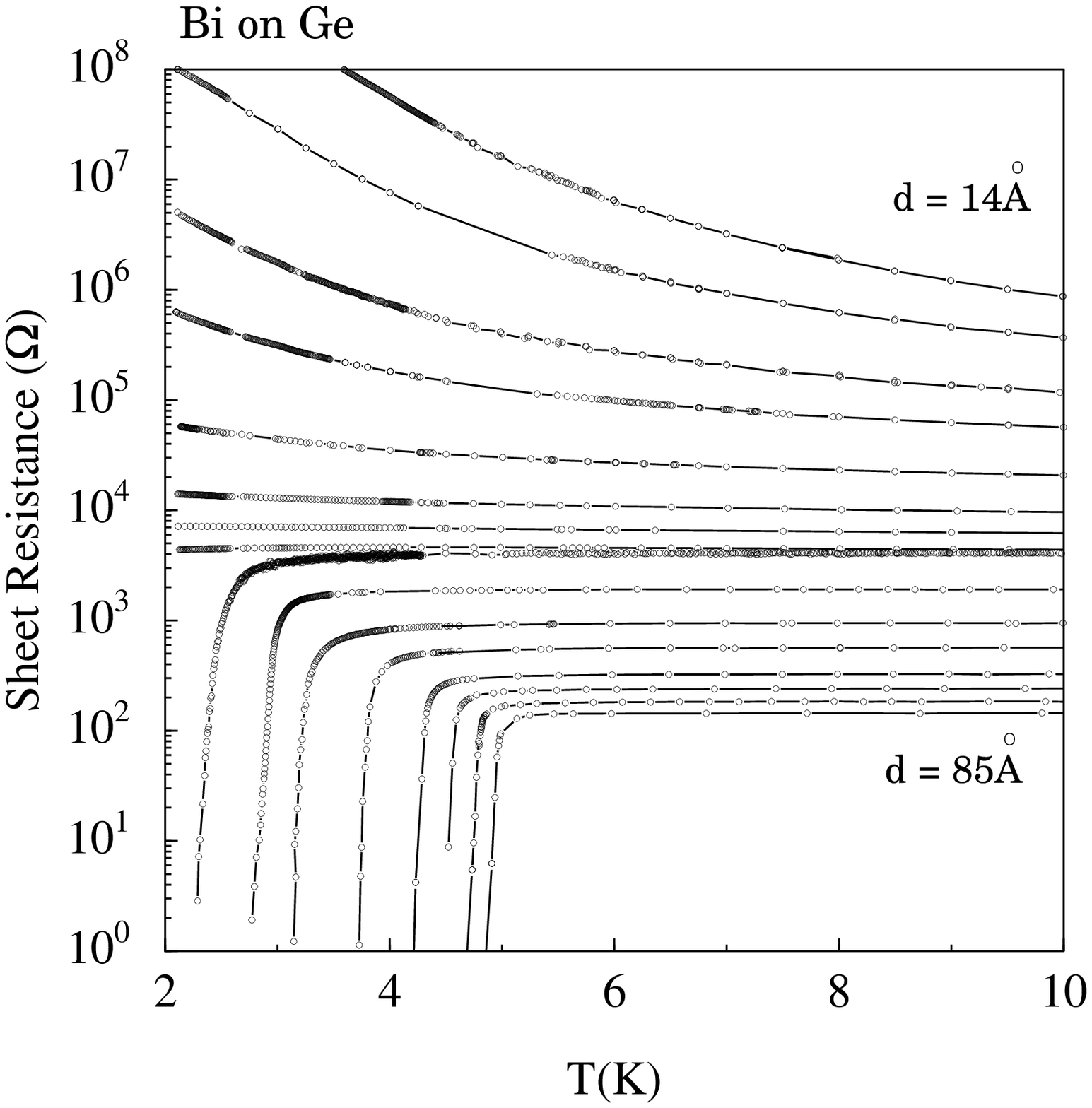}
\caption{ I-V characteristics for Bi films on bare quartz.}
\label{Fig. 1}\end{center}\end{figure}

\begin{figure}[b]
\begin{center}\leavevmode
\includegraphics[width=0.8\linewidth]{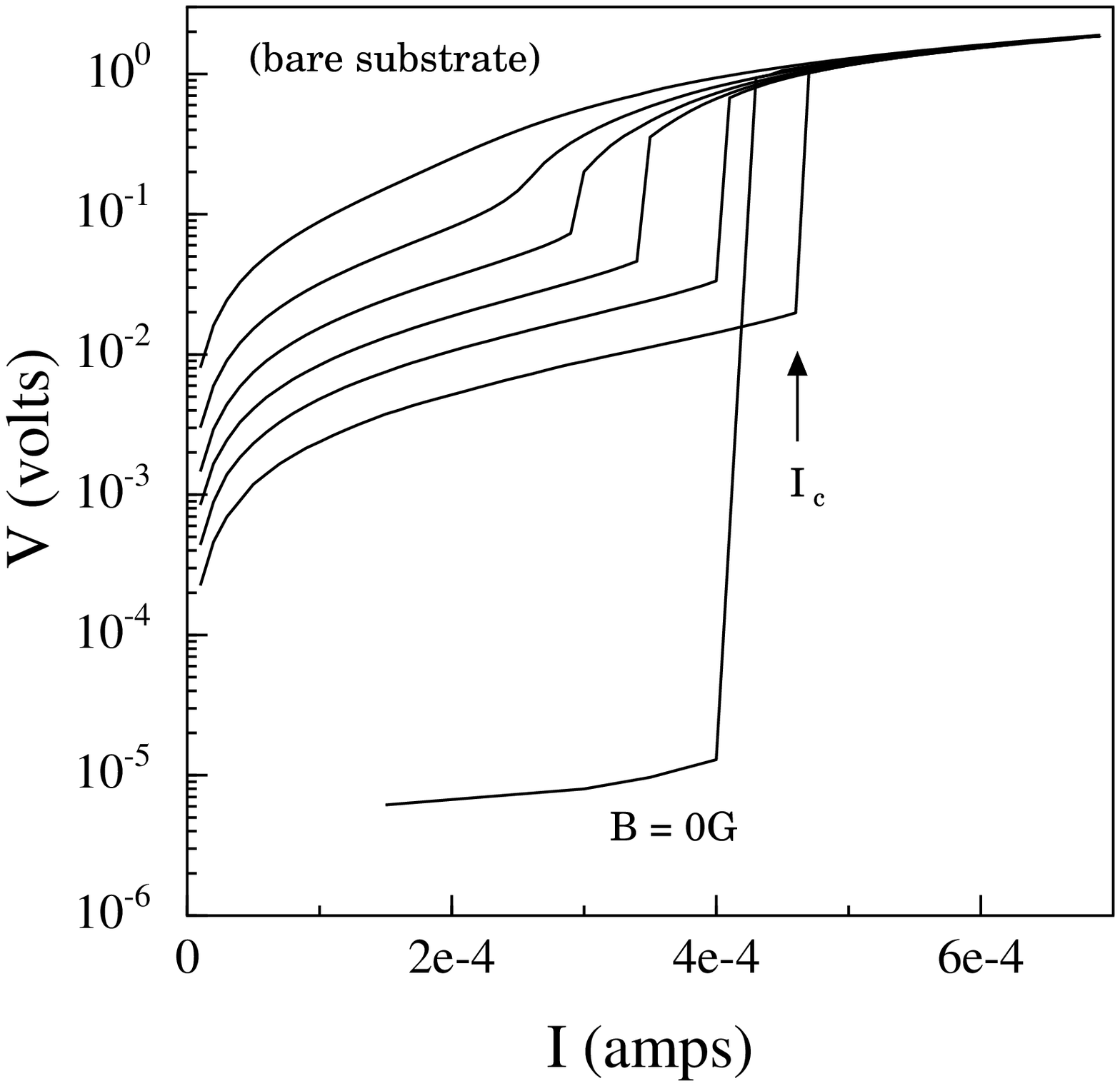}
\caption{ 
I-V characteristics for Bi films on Ge underlayers.
}\label{Fig. 2}\end{center}\end{figure}

A nonlinearity attributable to depinning seems to set in at lower
currents for the films on Ge. However, the $I_{\mathrm c}$ for films on Ge
is higher. In both these films, the dissipation is finite at any
applied field, as evidenced by a linear regime of the I-V,
from which a resistance can be discerned. This we attribute to thermally
activated vortex motion.

Fig. 3 shows the critical current as a function of normalized
temperature, at zero field for films on Ge as well as
on bare quartz. A peak in the critical current in the vicinity of
the transition temperature is clearly visible, for the films on
Ge. An identical phenomenon has been extensively investigated
in layered superconductors such as $2H-NbSe_{\mathrm 2}$~cite{shb}
and several other materials such as amorphous $Nb_{\mathrm 3}Ge$ and
$Mo_{\mathrm 3}Si$ films.~\cite{cct} This has been given the name ''peak effect.''
The pinning force density $F_{\mathrm p}$ exhibits a prominent peak
slightly below the upper critical field $B_{\mathrm c2}$ or near the critical
temperature. This effect is attributed to elastic instabilities
generated by local fluctuations of the pinning forces, which
induce a rapid softening of the flux line solid. This softening
allows the flux lines to conform readily to a configuration
that locks it to the inhomogeneities, so that the critical current
Ic increases.~\cite{shb,cct} So the observation of a peak in $I_{\mathrm c}$  implies
that an elastic flux line solid transforms to a plastic solid in
the vicinity of the peak. For the films on Ge, we identify the
first transition with the peak effect. In our
films, the critical fields are higher than what we can attain
using our magnet. Hence we have been unable to observe the
peak effect in the I-H plane. However, the observation in the
I-T plane is unambiguous proof of the peak effect.~\cite{shb,cct}

\begin{figure}[t]
\begin{center}\leavevmode
\includegraphics[width=0.8\linewidth]{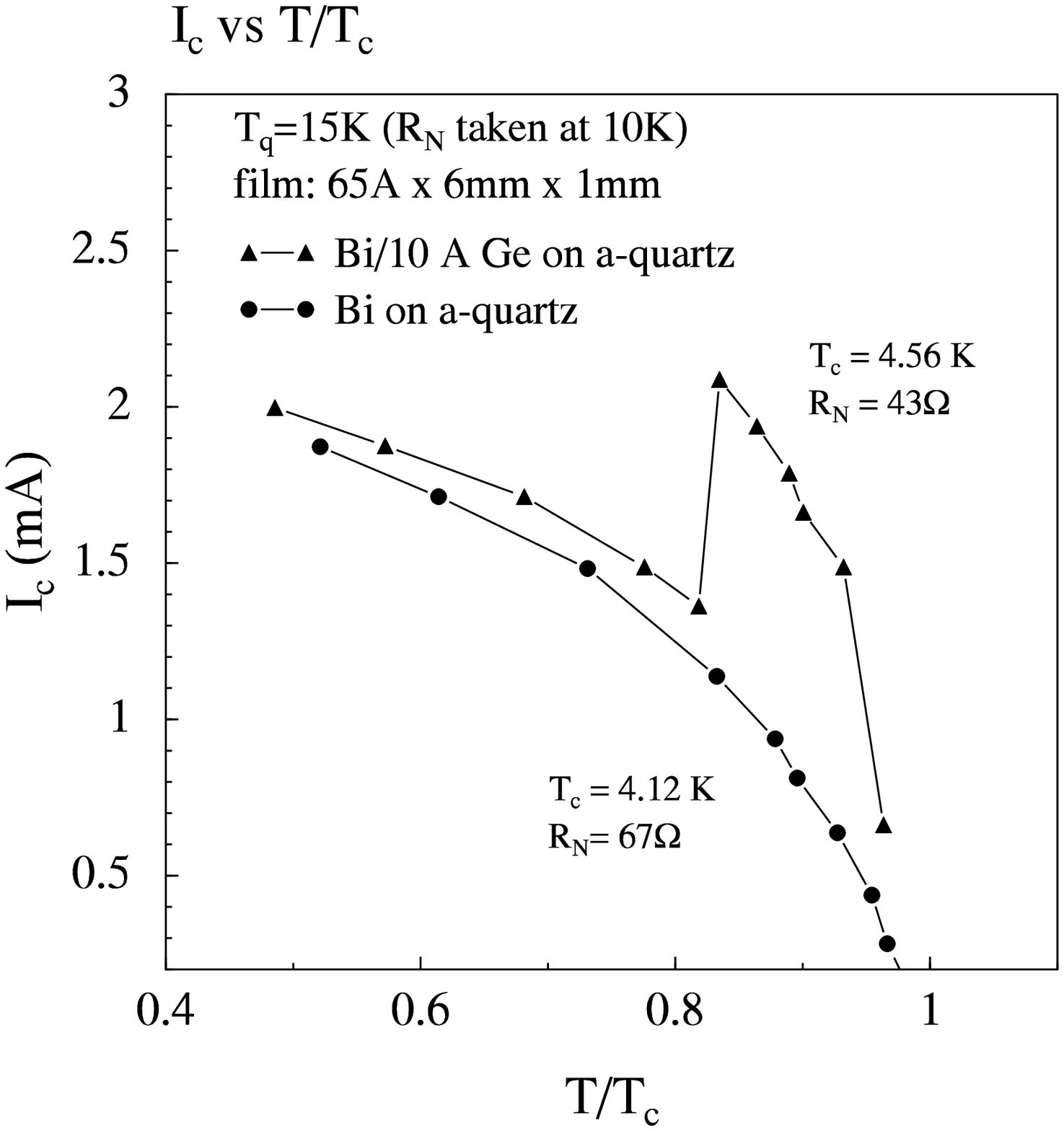}
\caption{ Variation of $I_{\mathrm c}$ with T.}
\label{Fig. 3}\end{center}\end{figure}

Clearly, there are interesting phenomena involving the vortex solid and liquid
phases in such ultrathin disordered films. Studies addressing the
potential landscapes in which the vortices move, and the details of
their motion are worth further investigations.

\begin{ack}
The work is supported by DST and UGC, Government of India. KDG thanks CSIR for the research fellowship.
\end{ack}

\end{document}